\documentclass{pasj00}
\draft                          

\begin{document}
\SetRunningHead{J.-M. Hur\'e}{``Dead zone'' in layered accretion models}
\Received{2002 March 5}
\Accepted{2002 August 7}

\title{Note on the ``Dead Zone'' in Layered Accretion Models}

\author{Jean-Marc \textsc{Hur\'e}}

\affil{LUTh/Observatoire de Paris-Meudon (FRE 2462 CNRS),\\ Place Jules
Janssen, 92195 Meudon Cedex, France\\ and\\ Universit\'e Paris 7 Denis Diderot,\\
2 Place Jussieu, 75251 Paris Cedex 05, France}

\email{jean-marc.hure@obspm.fr}
 
\KeyWords{accretion, accretion disks.}
 
\maketitle
              
\begin{abstract}                                                                Current layered accretion models neglect the properties of the ``dead zone''. However, as argued here from simple considerations, the thickness of this zone is a critical quantity when the disc is in hydrostatic equilibrium. It controls not only the structure of the superficial, active layers, but also the mid-plane density and the total disc mass, and should therefore be introduced in models  of that kind, steady or not. But in the absence of intrinsic heating, the dead zone must have a tiny size which, given the non-stationary and turbulent character of the global flow, makes very likely its mixing together with the two active layers.
\end{abstract}       

\section{Introduction}

Gammie (1996) proposed a two-phase accretion scenario for T Tauri discs to cure the expected inefficiency of the magneto-rotational instability (hereafter, MRI) in the cold, outer regions of the disc (beyond about $0.1$ AU from the central proto-star) due to the low abundance of electrons. In this model, only the superficial layers of the outer disc are made ``active'' thanks to incoming interstellar cosmic rays (that can reactivate the MRI), leaving a non-accreting ``dead zone'' around the equatorial plane. Interestingly, the so-called layered accretion disc model makes two major predictions: i) an infrared excess (a common feature in the spectrum of T Tauri stars; e.g. Bertout 1989) caused by a positive, radial gradient of the disc accretion rate, and ii) some ability to develop accretion bursts (the admitted interpretation for FU Orionis events; e.g. Hartman, Kenyon 1996; Kley, Lin 1999) due to mass accumulation in the outer regions. Some aspects of layered accretion have recently been investigated : evolution of the solar nebula (Stepinski 1999), occurrence of eruptive events through time-dependent simulations (Armitage et al. 2001), linear stability properties (Reyes-Ruiz 2001), possible applications to accretion in active galactic nuclei (Menou, Quataert 2001) and in binaries (Menou 2002), and the ionization of accreted material (Fromang et al. 2002). As outlined by these authors, there are still many uncertainties in such a toy-model. Apart from the hypothesis (we shall not discuss it here) that the source of angular momentum transport is only the MRI (see Stone et al. 2000, and references therein), the constraint imposed on the surface density $\Sigma_{\rm a}$ of the active layers (to be constant in space and time) appears as the strongest assumption. Misguidedly, the subsequent derivation of steady state profiles for the temperature, density, and disc thickness violates mass conservation (see also Menou 2002), or
\begin{equation}
4 \pi R \partial_t \Sigma_{\rm a} - \partial_R \dot{M}_{\rm a} \ne 0
\end{equation}
due to the non vanishing gradient of the total accretion rate, $\dot{M}_{\rm a}$. Another difficulty which is outlined here concerns the properties of the dead zone. None of the existing models have yet accounted for the structure of this zone. We argue here that it plays a crucial role. In particular, we demonstrate from simple arguments that it is not correct to find any reliable solution to the problem of layered accretion (steady or not) without specifying the thickness of the dead zone, either in a fully {\it ad-hoc} manner (i.e. by hand), or by considering explicitly its structure through coupled equations, as soon as the disc is in hydrostatic equilibrium. Finally, it turns out that this zone should be mixed due to its small vertical extent.

\section{Status and Importance of the Dead Zone in Layered Accretion}

\subsection{General Considerations}
\label{sec:gc}

\begin{figure}
  \begin{center}    
\FigureFile(160mm,160mm){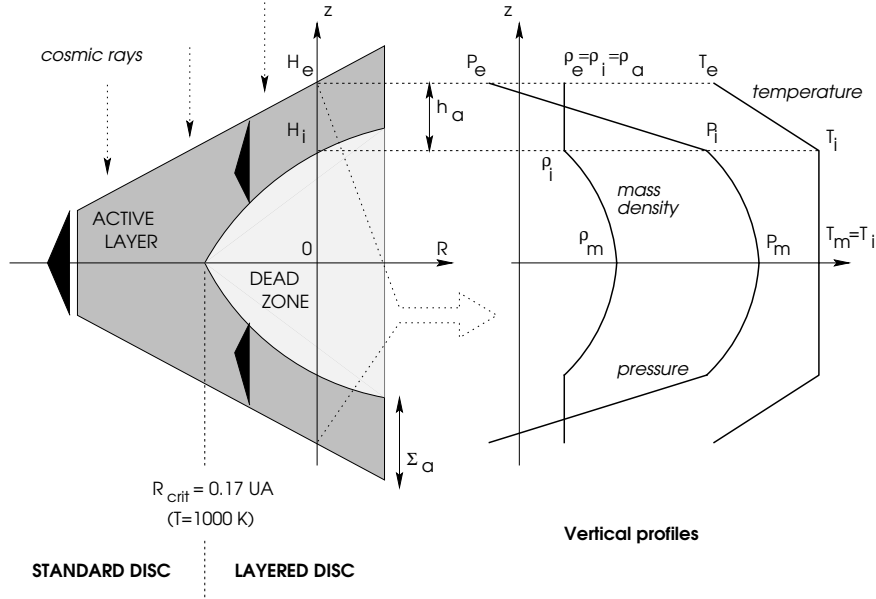}
  \end{center}    
  \caption{Schematic view (not to scale) of the layered accretion model (left) as originally proposed by Gammie (1996), and expected vertical profiles (right) for the mass density, pressure, and temperature in the dead zone and in the active layers (see subsection \ref{ss:dzone}).}
  \label{fig:lay}
\end{figure}

Let us start with the basic picture of the layered accretion model as originally described in Gammie (1996). In this model, the disc is divided into two parts, radially: an inner region assumed to be fully described by the standard theory of $\alpha$-discs (Shakura, Sunyaev 1973) and an outer, non-standard region. With respect to the central proto-star, the connection between the two regions is set by the temperature of gas and dust in the equatorial plane: below a critical value, $T_{\rm crit} \sim 10^3$K, thermal ionization alone is too weak, making inefficient the  magneto-rotational instability (MRI; see Stone et al. 2000) as the single source of angular-momentum transport. In the standard part of the disc, the mid-plane temperature decreases outwards so that the region with $T \le T_{\rm crit}$ is found at a radius larger than\footnote{In this paper, all quantities and computations are scaled to a one solar mass proto-star\label{note:sun}.  This choice has no noticeable consequences on the results exposed throughout the paper.} (e.g. Frank et al. 1992)
\begin{equation}
R_{\rm crit} \simeq 0.17 \, \kappa_{\rm R}^{2/9} \bar{\alpha}^{-2/9}  \dot{m}^{4/9} \qquad {\rm AU},
\end{equation}
where $\alpha = 0.01 \bar{\alpha}$ is the viscosity parameter (Shakura, Sunyaev 1973), $\kappa_{\rm R}$ is the Rosseland opacity of matter,\footnote{Near $10^3$ K, $\kappa_{\rm R} \approx 3-5$ cm$^2$ g$^{-1}$ with a very weak dependence on the density\label{note:kappa} (Pollack et al. 1994; Bell, Lin 1994).} and $\dot{m}$ is the mass accretion rate normalized to $10^{-8} M_\odot$ yr$^{-1}$, a typical value for accreting Class-II T Tauri stars (Calvet et al. 2000). In the layered disc, the MRI is maintained beyond $R_{\rm crit}$ with the help of cosmic rays, which can enhance the ionization degree at the disc surface, creating two superficial, accreting layers and an equatorial ``dead zone'' where gas and dust rotate without drifting inwards, as illustrated in figure \ref{fig:lay}. Assuming that only the MRI is present, the most critical assumption of this model is that the surface density of the active layers is $\Sigma_{\rm a} \approx 100$ g cm$^{-2}$ (for each one) for $R \ge R_{\rm crit}$.

\subsection{Basic Equations for the Active Layers. The Missing Closure Relation}

Let us now briefly recall the three basic equations for the disc beyond $R_{\rm crit}$. The first equation is dictated by energy balance. This equation links the effective temperature, $T_{\rm e}$, to the temperature $T_{\rm i}$ at the altitude $z=H_{\rm i}$ where  the dead zone joins the active layer (see figure \ref{fig:lay}). In the optically thick disc approximation, this is written as
\begin{equation}
T_{\rm i}^4 = \frac{3}{8} \kappa_{\rm R} \Sigma_{\rm a} T_{\rm e}^4.
\label{eq:energya}
\end{equation}

The second equation expresses the heating mode of the active layers. If all of the gravitationally energy release indeed occurs at $H_{\rm i}\le z \le H_{\rm i} + h_{\rm a}$ ($h_{\rm a}$ being the geometrical thickness of the active layer), the outgoing flux is
\begin{equation}
\sigma T_{\rm e}^4 = \frac{9}{4} \nu_{\rm a} \Sigma_{\rm a} \Omega_{\rm k}^2,
\label{eq:fluxa}
\end{equation}
where $\Omega_{\rm k}$ is the Keplerian rotation velocity, $\nu_{\rm a}=\alpha c_{\rm a} h_{\rm a}$ is the turbulent viscosity coefficient (Shakura, Sunyaev 1973) and $c_{\rm a}$ is the sound speed. If the ionization rate has its interstellar medium value, cosmic rays do not contribute to the local disc heating (e.g. D'Alessio et al. 1998).

The third equation follows from the hydrostatic equilibrium of {\it the whole vertical structure}. Current models neglect the dead zone at this level, but this is not correct: two equations must be considered with the condition of pressure balance at $z=H_{\rm i}$. Un-balancing the pressure would create a mean flow in the direction of the pressure gradient (i.e. vertically here). With the usual, vertically averaged approach and neglecting disc self-gravity (see subsection \ref{sec:dsg}), we find for the active layer
\begin{equation}
\frac{P_{\rm i}}{\rho_{\rm i}} = \Omega_{\rm k}^2 h_{\rm a} \left( H_{\rm i} + h_{\rm a} \right) =  \frac{k T_{\rm i} }{\mu_{\rm a} m_{\rm H}},
\label{eq:hea}
\end{equation}
where $\mu_{\rm a} m_{\rm H}$ is the mean mass per particle (with $\mu_{\rm a} \approx 2.37$ for a molecular gas with cosmic abundances) and $m_{\rm H}$ is the proton mass. A very similar equation can be derived by considering vertical stratification, for instance from a polytrope. This relation shows the coupling between the active layers and the dead zone. The point is that, physically, the thickness of the dead zone determines the strength of vertical gravity at $z \ge H_{\rm i}$, and therefore the structure (i.e. temperature, density $\rho_{\rm a}$, and thickness $h_{\rm a}=\frac{\Sigma_{\rm a}}{\rho_{\rm a}}$) of the accreting layers. Note that, with an appropriate formula for the Rosseland opacity, one recovers Gammie's (unsteady) solution from equations (\ref{eq:energya}), (\ref{eq:fluxa}), and (\ref{eq:hea}) by setting $H_{\rm i} \sim 0$ (or $H_{\rm i} \ll h_{\rm a}$), which ultimately means that there is almost {\it no dead zone}.

\subsection{Structure of the Dead Zone}
\label{ss:dzone}

In the spirit of the layered disc model, there is not heat generation at $|z| \le H_{\rm i}$ meaning that the dead zone is necessarily isothermal, vertically. So, the mid-plane temperature is $T_{\rm m}=T_{\rm i}$ and the pressure and density obey a Gaussian distribution (assuming a constant chemical content). Actually, hydrostatic equilibrium of the dead zone yields
\begin{equation}
\rho_{\rm m} = \rho_{\rm i} \exp \left( \frac{H_{\rm i}}{\lambda}\right)^2,
\label{eq:gaussian}
\end{equation}
where $\lambda$ is given by [see equation (\ref{eq:hea})]
\begin{equation}
\lambda^2 = \frac{2 k T_{\rm m} }{\mu_{\rm m} m_{\rm H} \Omega_{\rm k}^2} \approx 2 h_{\rm a}  \left( H_{\rm i} + h_{\rm a} \right)
\end{equation}
with $\mu_{\rm m} \approx \mu_{\rm a}$ (this should be correct given temperatures of interest). The surface density of the dead zone $\Sigma_{\rm d}$ is then given by
\begin{equation}
\Sigma_{\rm d} =  2 \int_0^{H_{\rm i}}{\rho_{\rm m} \exp \left( - \frac{z^2}{\lambda^2}\right) dz} = \sqrt{\pi}\rho_{\rm m} \lambda \; {\rm erf}\left(\frac{H_{\rm i}}{\lambda} \right),
\label{eq:sigmad}
\end{equation}
where ${\rm erf}$ denotes the Error function. It follows that a large-size dead zone (i.e. $H_{\rm i} \gg h_{\rm a}$) can lead to large mid-plane densities and, thus, to a large-mass disc. In other words, the dead zone must have a very small vertical extent compared to the active layers. This even raises the question of the existence of such a zone: because it is squeezed between two turbulent layers, it would probably mix with them and tend to disappear in a turn-over time scale. We shall now examine to what extent the structure of the active layers is sensitive to the presence of the dead zone.

\section{Essential Role of the Dead Zone}

\subsection{The Strong Coupling}

We can see that equations (\ref{eq:energya}), (\ref{eq:fluxa}), and (\ref{eq:hea}) involve four unknowns ($T_{\rm i}$, $T_{\rm e}$, $h_{\rm a}$, and $H_{\rm i}$), leaving the system under-determined. As shown above, $H_{\rm i}$ is still unconstrained, but must be specified in some fashion. This is true whatever the state of the disc, {\it steady or unsteady}. In current models, the choice is $H_{\rm i}=0$, as already mentioned. An interesting question is then: how is the structure of the active layers modified if the parameter $H_{\rm i}$ is set to another value? In other words, what is the sensitivity of this special choice on the topology of layered solutions. To answer this question, we can simply combine equations (\ref{eq:energya}), (\ref{eq:fluxa}), and (\ref{eq:hea}) to get
\begin{equation}
h^5_{\rm a} (h_{\rm a}+H_{\rm i})^7 = \frac{729 \, k^8 \kappa_{\rm R}^2 \alpha^2 \Sigma_{\rm a}^4}{1024 \, \sigma^2 \mu_{\rm a}^8 m_{\rm H}^8 \Omega_{\rm k}^{10}},
\label{eq:hihaanalytic}
\end{equation}
which is a constant at a given radius (see footnote \ref{note:kappa}). It is therefore clear from that relation that $H_{\rm i}$ and $h_{\rm a}$ cannot be specified independently. If $h_{\rm a} \gg H_{\rm i}$, one tends to the no-dead zone solution and, in the reverse case, the coupling is extreme, since $h^5_{\rm a} H^7_{\rm i} \sim cst$.

We can also answer the above question by considering the steady state hypothesis, which provides a natural closure relation for our system of four equations.  In some sense, this is more satisfactory than imposing in equation (\ref{eq:hihaanalytic}) some arbitrary values for $H_{\rm i}$, which, for instance, is  done in Stepinski (1999). As we shall see, this hypothesis leads to an unrealistic disc, but this is unimportant here, since we only wish to exhibit, from a mathematical point of view, the sensitivity of the structure of the layered disc to $H_{\rm i}$ when a physical closure relation is introduced. Thus, if the flow is steady, we have (e.g. Frank et al. 1992)
\begin{equation}
\frac{1}{2}\dot{M}_{\rm a} = 3 \pi \nu_{\rm a} \Sigma_{\rm a} = cst,
\end{equation}
where $\dot{M}_{\rm a}$ is the accretion rate for the two active layers. Because $\Sigma_{\rm a}=cst$ in the present problem (see subsection \ref{sec:gc}), we must also have $\nu_{\rm a}=cst$. This is possible with the viscosity prescription if $c_{\rm a}h_a=cst$ or
\begin{equation}
T_{\rm i} h_{\rm a}^2 = cst = \frac{ \mu_{\rm a} m_{\rm H} \dot{M}_{\rm a}^2 }{36 \pi^2 \alpha^2 k \Sigma_{\rm a}^2} \simeq 0.14 \, \dot{m}^2 \bar{\alpha}^{-2} \qquad {\rm AU}^2 \; {\rm K}.
\label{eq:vis}
\end{equation}
This is the missing fourth relation.
  
\subsection{Discussion of Steady State Solutions}
\label{sec:discuss}

With the assumption of a steady state, the structure of the active layers and that of the dead zone can be fully determined in a self-consistent way. From equations (\ref{eq:energya}), (\ref{eq:fluxa}), and (\ref{eq:vis}), we find that each active layer has a thickness (see footnote \ref{note:sun}) of
\begin{equation}
h_{\rm a} \simeq 2.6 \times 10^{-2} \, \kappa_{\rm R}^{-1/8} \bar{\alpha} \dot{m}^{7/8} x^{3/8} \qquad {\rm AU},
\label{eq:ha}
\end{equation}
where $x = \frac{R}{1 \, \rm AU}$ and the actual value of the opacity coefficient is fixed by the local temperature and pressure. However, the point is that $h_{\rm a}$ must also satisfy equation (\ref{eq:hea}) with the condition that $H_{\rm i} > 0$, which is not guaranteed. Actually, equation (\ref{eq:hea}) gives
\begin{equation}
H_{\rm i}  + h_{\rm a} \simeq 3.2 \times 10^{-2} \, \kappa_{\rm R}^{3/8} \bar{\alpha}^{-1} \dot{m}^{-5/8}  x^{15/8} \qquad {\rm AU},
\label{eq:haplushi}
\end{equation}
and it follows from equations (\ref{eq:ha}) and (\ref{eq:haplushi}) that $H_{\rm i}$ vanishes for
\begin{equation}
R_{\rm in} \simeq 0.87 \,  \kappa_{\rm R}^{-1/3} \bar{\alpha}^{-4/3}  \dot{m} \qquad {\rm AU},
\end{equation}
meaning that the steady solution exists at $R \ge R_{\rm in}$ only. Generally, we have $R_{\rm crit} \ne R_{\rm in}$. Even $R_{\rm crit} \ll R_{\rm in}$ for usual parameter values, implying that the outer layered solution does not connect smoothly to the inner standard disc solution. The reason is simple: the conditions $\int_0^\infty{\rho dz}=\Sigma_{\rm a}$ and $T=T_{\rm crit}$ are never satisfied simultaneously (i.e. at the same radius) in an $\alpha$-disc, except if precisely
\begin{equation}
\dot{m} \simeq 5.4 \times 10^{-2} \, \bar{\alpha}^2 \kappa_{\rm R}(T_{\rm crit}) \approx  0.2 \, \bar{\alpha}^2 \equiv \dot{m}_0,
\label{eq:mdot0}
\end{equation}
which is very restrictive. This smooth connection might, however, exist by relaxing the value of $T_{\rm crit}$ or $\Sigma_{\rm a}$, for instance, or by invoking some variation of the $\alpha$-parameter value with the radius, which is very speculative.

\begin{figure}
  \begin{center}    
\FigureFile(160m,160m){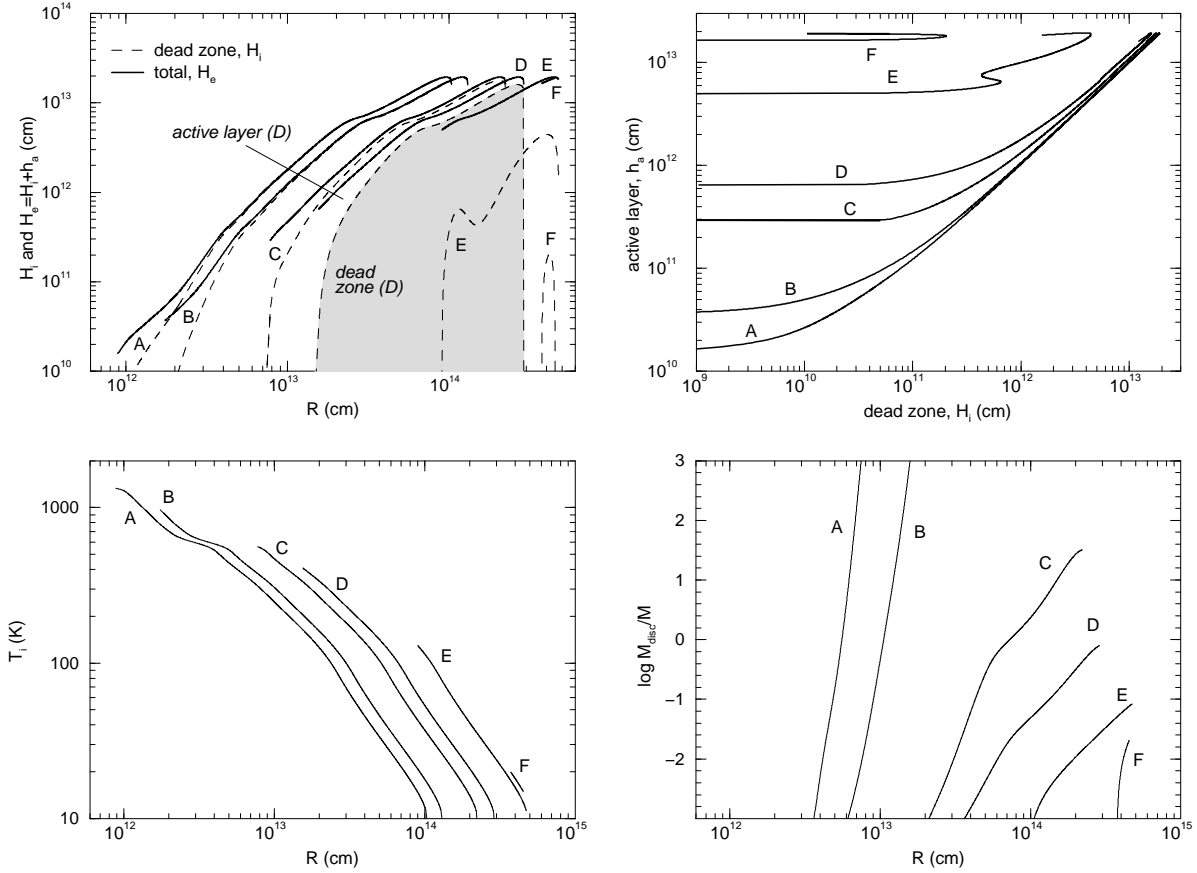}
 \end{center}  
  \caption{Vertically averaged solutions for the steady, layered accretion disc model (see footnote \ref{note:sun}) computed for $\alpha=0.01$ and various accretion rates:  $\dot{m}=0.1$ (A),  $\dot{m}=0.2=\dot{m}_0$ [B; in this case, equation (\ref{eq:mdot0}) is exactly fulfilled, meaning that there is a continuous connection between the inner standard disc and the layered disc], $\dot{m}=1$ (C), $\dot{m}=2.2$ (D), $\dot{m}=10$ (E), and $\dot{m}=13$ [F; see also equation (\ref{eq:mdotlim})]. The disc mass, $M_{\rm disc}$, only accounts for the matter (dead zone and active layers) at $R \ge R_{\rm in}$.}
  \label{fig:lay1D}
\end{figure}

Thus, for $\dot{m} > \dot{m}_0$, there is no steady solution between $R_{\rm crit}$ and $R_{\rm in}$. This no-solution domain can be quite wide, especially for large accretion rates (as expected in very young systems) and/or for large $\alpha$-parameter values. According to equations (\ref{eq:ha}) and (\ref{eq:haplushi}), large accretion rates imply large-size active layers and a small-size dead zone, whereas low accretion rates mean small-size active layers and a prominent, and possibly massive, dead zone. Conversely, for $\dot{m} < \dot{m}_0$, the total surface density at $R_{\rm crit}$ is less than $2 \Sigma_{\rm a}$; the inner standard disc and the outer layered disc overlap between $R_{\rm in}$ and $R_{\rm crit}$. In this case, an acceptable solution smoothly connecting the two discs should be easily found by changing the equation for the inner disc (this exercise is out of the scope of this paper; see subsection \ref{sec:dsg}). But this concerns quite evolved (WTTS-like) systems with accretion rates lower than $\sim 2 \times 10^{-9} M_\odot$ yr$^{-1}$ (e.g. Calvet et al. 2000), and the amount of gas available in these discs has strongly decreased (see subsection \ref{sec:dsg}).

These quantitative results have been fully confirmed by numerical computation of the solutions\footnote{For this exercise, we have used realistic Rosseland opacities as compiled in Hur\'e (2000).} of equations (\ref{eq:energya}), (\ref{eq:fluxa}), (\ref{eq:hea}), and (\ref{eq:vis}). The results obtained for $\alpha=0.01$ and six different accretion rates are displayed in figure \ref{fig:lay1D}. There is no solution for accretion rates larger than\footnote{This limit is in fact compatible with the value deduced analytically by equating $R_{\rm in}$ and $R_{\rm out}$, the marginally optically thin, outer disc radius where $T_{\rm e} \approx T_{\rm i} \simeq 10$ K, namely $R_{\rm out} \simeq 17 \, \dot{m}^{1/3} {\rm AU}$. At $10$ K, $\kappa_{\rm R} \simeq 0.02$ cm$^2$ g$^{-1}$ (see references given in footnote 2) and the optical thickness of each active layer is $\sim 2$, in the Rosseland sense.} 
\begin{equation}
\dot{m}_{\rm lim} \approx 13 \, \bar{\alpha}^2,
\label{eq:mdotlim}
\end{equation}
and the sensitivity of the solution to the accretion rate and viscosity parameter value is very important (see the morphological change from runs C to F, whereas the accretion rate is increased by only a factor $6$). As already announced, the steady solution appears to be non-physical either because it leaves a gap (the no-solution domain; see for instance run E where $R_{\rm crit} \approx 0.6$ AU and $R_{\rm in } \approx 6$ AU where $T_{\rm i} \sim 130$ K), or because it involves discs more massive than the central object by orders of magnitudes (see runs C, B, and A), which is not plausible. In any case, the most important point that these runs demonstrate is that the structure of the active layers is strongly sensitive to the structure of the dead zone, that is to $H_{\rm i}$, which, indeed, can not be ignored. It is therefore expected that this sensitivity also be present in time-dependent models, as soon as hydrostatic equilibrium is assumed.

\subsection{Disc Self-Gravity}
\label{sec:dsg}

The mass of gas contained in the active layers is
\begin{equation}
M_{\rm al} = 4 \pi \Sigma_{\rm a} \int_{R_{\rm crit} \ll R}^R{R'dR'}\sim 7 \times 10^{-5} \left( \frac{R}{{\rm AU}} \right)^2 M_\odot.
\label{eq:dmass}
\end{equation}

Therefore, this is a firm, lower limit for the mass of the whole disc: for instance, for $H_{\rm i}/h_{\rm a}=1$, the disc mass would be $\sim 4 \times M_{\rm al}$. According to Gammie's no-dead zone solution, hydrostatic equilibrium is expected to be modified by self-gravity as close as $\sim 20$ AU (the distance where $2\frac{M_{\rm al}}{M} \gtrsim \frac{h_{\rm a}}{R}$) from the center and beyond. In this (vertically and probably gravitationally unstable) self-gravitating regime, it is possible to show again the influence of the dead zone on the structure of the active layers using the infinite-slab approximation (Paczy\'nski 1978). Actually, equation (\ref{eq:hea}) would then become 
\begin{equation}
\frac{P_{\rm i}}{\rho_{\rm i}} =  h_{\rm a} \left[\Omega_{\rm k}^2 (h_{\rm a}+H_{\rm i}) + 4 \pi G \left( \Sigma_{\rm a} + \frac{1}{2}\Sigma_{\rm d} \right) \right].
\label{eq:hea-withsg}
\end{equation}

In particular, in the pure self-gravitating limit, this equation becomes
\begin{equation}
\frac{P_{\rm i}}{\rho_{\rm i}} = 4 \pi G h_{\rm a} \Sigma_{\rm a} \left( 1 + \frac{\Sigma_{\rm d}}{2 \Sigma_{\rm a}}  \right).
\label{eq:hea-puresg}
\end{equation}

The second term in the brackets vanishes when $H_{\rm i} \ll h_{\rm a}$ (this is no-dead zone case). The point is that the temperature in active layers from equation (\ref{eq:hea-puresg}) is strongly sensitive to $H_{\rm i}$. This means that the dead zone has, again, a major influence on the structure of the active layers, and that the problem with massive discs invoked above remains.

\section{Concluding Remarks}

Despite its ``passive'' role with respect to the assumed mechanism of angular-momentum transport and heat generation, the dead zone is an essential component of layered accretion-disc models. Whatever the importance of self-gravity, this zone mechanically supports the active layers, and thus determines the global properties of the disc as a whole. Any layered model of that kind, steady or not, where the disc has reached hydrostatic equilibrium (see also Glassgold et al. 2000), must take the dead zone into account, and in particular its thickness, at least in the form of a free parameter. Probably, a model accounting for more physical mechanisms (and containing more degrees of freedom) could modify some issues raised here, and even might produce steady state solutions. However, it appears clearly that, in the absence of internal heating, the dead zone must have a small thickness with respect to the active layers; otherwise, the disc would be extremely massive. Also, without any noticeable extent, the dead zone should not survive between the two turbulent layers, but mix with them. Besides, observations of discs around young stars indicate that they have a mass not in excess of the central mass (e.g. Calvet et al. 2000) as well as an outward decrease of the surface density (Beckwith et al. 1990; Dutrey et al. 1998), both properties of which are not compatible with the layered accretion model.\\

It is a pleasure to thank my colleagues S. Collin, D. Gautier, F. Hersant, D. Richard, and J.-P. Zahn for stimulating discussions. I am grateful to the referee for important suggestions and comments to improve the paper.

\end{document}